\documentclass[aps,prb,twocolumn,superscriptaddress,floatfix]{revtex4-2}
\pdfoutput=1 
\usepackage[colorlinks=true,citecolor=blue,linkcolor=blue,breaklinks=true]{hyperref}
\usepackage{amssymb} 
\usepackage{amsmath} 
\usepackage{physics} 
\usepackage{booktabs}
\usepackage{times} 
\usepackage{setspace} 
\usepackage{xfrac} 
\usepackage{placeins} 
\usepackage{multirow}
\usepackage{array}
\usepackage{siunitx}

\usepackage{xcolor}

\newcommand{\NC}{NdCl$_3$ }
\newcommand{\NCns}{NdCl$_3$}

\begin{document}


\title{Frustration induced dimensional reduction and coexistence of long and short-range magnetic order in NdCl$_{3}$}
\thanks{This manuscript has been authored by UT-Battelle, LLC under Contract No. DE-AC05-00OR22725 with the U.S. Department of Energy.  The United States Government retains and the publisher, by accepting the article for publication, acknowledges that the United States Government retains a non-exclusive, paid-up, irrevocable, world-wide license to publish or reproduce the published form of this manuscript, or allow others to do so, for United States Government purposes.  The Department of Energy will provide public access to these results of federally sponsored research in accordance with the DOE Public Access Plan (http://energy.gov/downloads/doe-public-access-plan)}


\author{Eli Zoghlin}
\email[]{zoghlinea@ornl.gov}
\affiliation{Materials Science \& Technology Division, Oak Ridge National Laboratory, Oak Ridge, Tennessee 37831, USA.}

\author{Matthew B. Stone}
\author{Vasile O. Garlea}
\author{Matthias D. Frontzek}
\affiliation{Neutron Scattering Division, Oak Ridge National Laboratory, Oak Ridge, Tennessee 37831, USA}
\author{Andrew D. Christianson}
\affiliation{Materials Science \& Technology Division, Oak Ridge National Laboratory, Oak Ridge, Tennessee 37831, USA.}
\author{Andrew F. May}
\email[]{mayaf@ornl.gov}
\affiliation{Materials Science \& Technology Division, Oak Ridge National Laboratory, Oak Ridge, Tennessee 37831, USA.}

\date{\today}

\begin{abstract}
The phenomenon of frustration greatly enriches the accessible physics of quantum magnets. With this in mind we study the magnetism of \NC using a combination of bulk properties measurements and neutron scattering techniques. The low-temperature heat capacity reveals two magnetic transitions at $T_{N1}$ = 270 mK and $T_{N2}$ = 180 mK. However, much of the magnetic entropy is released above $T_{N1}$, manifesting as a broad peak centered at $T^{*} \approx$ 450 mK. A similar broad peak is observed in the magnetic susceptibility, which possesses a clear anisotropy; this data can be well fit using a 1-dimensional XXZ model, yielding an antiferromagnetic nearest-neighbor exchange $J_{NN}$ = -0.962(2) K and exchange anisotropy close to the Ising limit. Single crystal elastic neutron scattering reveals highly anisotropic magnetic diffuse scattering above $T_{N1}$, confirming that quasi-one-dimensional, short-range, antiferromagnetic order is the origin of the broad peak in the heat capacity and magnetic susceptibility. Magnetic Bragg peaks characterized by a $\vec{k}$ = ($0$ $0$ $\frac{1}{2}$) propagation vector emerge below $T_{N1}$. Interestingly, the magnetic diffuse scattering persists for $T_{N2} < T < T_{N1}$, indicating a regime of coexisting short and long-range order. The magnetic Bragg peaks exhibit an additional increase in intensity below $T_{N2}$ with no change in $\vec{k}$. Concomitantly, the diffuse scattering disappears indicating the attainment of full long-range order. While the precise nature of the ordered magnetic ground state remains unresolved, the observed magnetic scattering indicates predominate $c$-axis moments with a small $ab$-plane component. We propose that the quasi-one-dimensional behavior and the coexistence of short and long-range order are driven by frustration of exchange interactions with significant anisotropy.
\end{abstract}

\pacs{}

\maketitle

\section{Introduction \label{intro}}

Magnetic frustration is one key mechanism for realizing novel magnetic phases possessing complex spin structures and dynamics. In frustrated models, a combination of the lattice topology and the operative exchange interactions creates a strong competition between degenerate ground states \cite{rau2019frustrated}. For some models, regions of the phase diagram may evade long-range order entirely, with quantum fluctuations stabilizing a magnetically disordered ground state such as the long-heralded quantum spin liquid (QSL) \cite{savary2017quantum,knolle2019field}. In real materials, magnetic order typically occurs, but at a temperature significantly below the energy scale of the exchange interactions. Additional frustration induced effects may also manifest in the nature of the magnetic order and the process by which it occurs. 

One such effect is the reduction of the dimensionality of the spin degree of freedom from what is expected by a simple inspection of the nuclear structure. This occurs, for example, in SrCu$_{2}$O$_{3}$, where quasi-one-dimensional (quasi-1D) spin ladders are stabilized out of a nuclear structure with dominant 2D character \cite{azuma1994observation}. A similar reduction from 2D to 1D has been observed in several anisotropic triangular lattice materials, such as  $\alpha$-NaMnO$_{2}$ \cite{dally2018amplitude} and Ca$_{3}$ReO$_{5}$Cl$_{2}$ \cite{hirai2019one}. Frustration induced dimensional modification has also been observed in materials with fully 3D nuclear structures, such as the emergence of quasi-1D behavior in the spinel ZnV$_{2}$O$_{4}$ \cite{lee2004orbital} and quasi-2D behavior in the isotropic tetrahedral cluster material pharmacosiderite, (H$_{3}$O)Fe$_{4}$(AsO$_{4}$)$_{3}$(OH)$_{4}\cdot$5H$_{2}$O \cite{okuma2021dimensional}.

In order to further study these kinds of modifications, it is desirable to identify material families on a frustrated lattice that can be tuned by isostructural substitution of the magnetic ion. Frustrated rare-earth-based material families are compelling in this regard. The relatively strong spin-orbit coupling (SOC) often enhances the presence of anisotropic exchange interactions, which can lead to extremely rich phase diagrams \cite{rau2019frustrated, li2016anisotropic,yan2017theory}. Importantly, selecting different rare-earth ions changes the nature of the crystal electric field (CEF) ground state and its attendant anisotropies. This allows for electronic tunability while preserving the frustrated lattice topology. Furthermore, the CEF may stabilize an effective $J_{eff} = \frac{1}{2}$ ground state, potentially enhancing quantum fluctuations and enriching frustration-driven physics. With these factors in mind we have turned to the binary rare-earth trichlorides, $RE$Cl$_{3}$, ($RE$ = rare-earth ion), which offer a chemically simple experimental platform for such studies. 

\begin{figure}
 \includegraphics[width =8.6 cm]{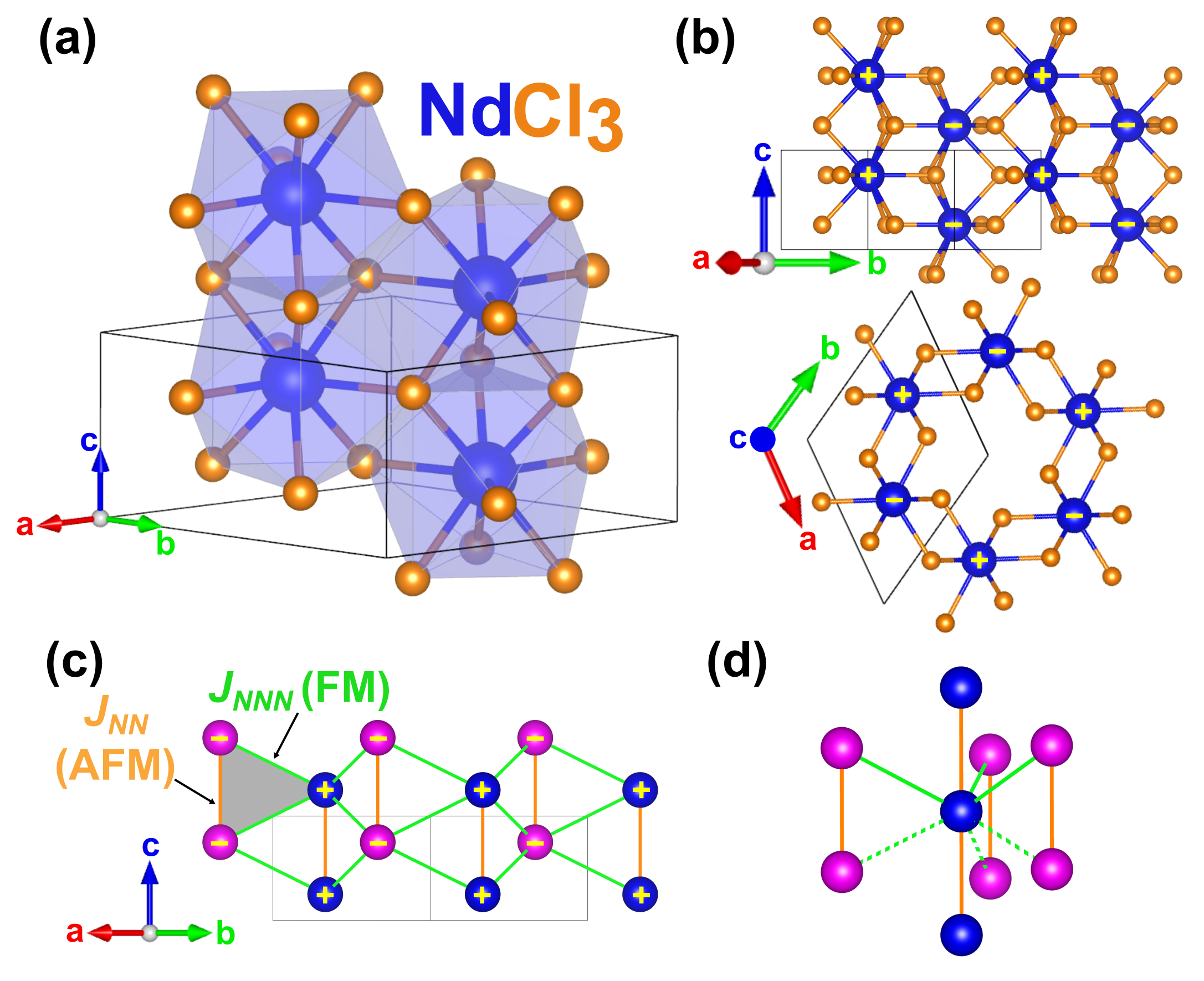}
 \caption{\label{fig:NuclearStructure} (a) Illustration of the crystal structure of \NC refined using the $T$ = 45 mK elastic neutron scattering data in space group P6$_{3}$/m (\#176). The black solid line represents the hexagonal unit cell. The transparent blue polyhedra show the trigonal coordination of Nd by Cl. (b) Projection views of the structure along the [2 1 0] and [0 0 1] lattice directions (top and bottom, respectively). The yellow ``+'' and ``-'' indicate the relative shifts of Nd ions into or out of the page. (c) Illustration of the NN ($J_{NN}$, orange) and NNN ($J_{NNN}$, green) exchange bonds viewed along the [1 1 0] lattice direction. Only the Nd ions are shown and the different coloring represents the bipartite shifted chains, as discussed in the main text. The gray filled-in region highlights the isosceles triangle of which the exchange network is composed. (d) Restricted view showing $J_{NNN}$ for only one Nd ion. The $J_{NNN}$ are divided into subsets (solid and dashed green lines). For dominant AFM spin correlations along the chains it is not possible to fully satisfy both subsets.}
 \end{figure}

In this family, $RE$ = La -- Gd form in the hexagonal space group P6$_{3}$/m (\#176) while various structures are observed for Tb and heavier rare-earth elements. In the hexagonal compounds, the $RE$ ions occupy a symmetric trigonal prismatic coordination of Cl$^{-}$ (site symmetry $C_{3h}$). These $RE$ polyhedra form a puckered arrangement in the $ab$-plane, with every other $RE$ ion displaced along the $c$-axis (see Fig. \ref{fig:NuclearStructure}). This leads to a 3D structure built from $RE$-Cl-$RE$ bonding pathways. Additionally, the $RE$-Cl polyhedra form face-sharing chains along the \textit{c}-axis, which are connected to neighboring chains via edge-sharing connections. 

The single-ion physics of the hexagonal compounds has been studied using dilute (La$_{1-x}RE_{x}$)Cl$_{3}$ samples, providing detailed analyses of the CEF level schemes and $g$-factors via optical techniques and electron spin resonance (ESR), respectively \cite{judd1959analysis,carlson1961state,eisenstein1963spectrum,hutchison1958paramagnetic,schmid1987structure,prinz1966optical,crosswhite1976spectrum}. Both properties vary significantly across the series. For \NCns, optical measurements show that the first excited CEF state is found at $E \approx$ 14 meV (163 K), providing a well-isolated doublet at low temperature \cite{carlson1961state,eisenstein1963spectrum}. ESR measurements at $T \approx$ 4 K show a uniaxial $g$-factor anisotropy with $g_{||}$ = 3.996(1) and $g_{\perp}$ = 1.763(1), referenced to the $c$-axis \cite{hutchison1958paramagnetic}. 

Relative to the heavier $RE$ compounds \cite{murasik1985magnetic,kramer1999noncollinear,dieke1961absorption,sala2019crystal,sala2021van,sala2023field} less information is available on the microscopic details of the magnetism in the hexagonal compounds. Magnetic susceptibility and heat capacity measurements have been conducted across the series and these data demonstrate that GdCl$_{3}$ posses ferromagnetic (FM) ordering at $T$ = 2.2 K \cite{wolf1962ferromagnetism}, while lower temperature ($T <$ 1 K) AFM behavior is found otherwise. In CeCl$_{3}$ there is evidence of multiple magnetic transitions below $T$ = 200 mK \cite{colwell1969low,colwell1967low, cothrine2025thesis,eisenstein1965low,landau1973thermal}. Notably, no direct evidence of long-range magnetic order in SmCl$_{3}$ and \NC has yet been provided \cite{colwell1967low,colwell1969low,eisenstein1965low} \footnote{Several reports of the magnetic susceptibility for \NC show peaks at temperatures above $T$ = 1 K which appear only weakly in corresponding heat capacity measurements \cite{eisenstein1965low,colwell1969low,colwell1967low}. They are most likely due to unresolved impurities in the previously studied samples. No such peaks were observed in our susceptibility measurements}. A broad maximum in the heat capacity and magnetic susceptibility of \NC has been observed at $T\approx$ 450 mK, which has been ascribed to quasi-1D short-range order based on its functional form \cite{colwell1969low,colwell1967low,eisenstein1965low}. Comparison of this broad peak to the AFM Ising chain model yielded an estimate of the nearest-neighbor (NN) exchange interaction $|J_{NN}| \approx$ 1 K \cite{colwell1969low}. Finally, antiferroelectric order below $T$ = 420 mK has been identified in PrCl$_{3}$ \cite{harrison1976one,hessler1971low, colwell1969low}. 

More direct insight into the nature of the spin-spin interactions has been primarily gained from analysis of ESR data \cite{birgeneau1967magnetic,baker1966determination,baker1968high,brower1966electron, riley1970interactions}. Analysis of this ESR data for \NC established that the $g$-factor anisotropy generates an Ising-like ($XXZ$-type) exchange anisotropy and was used to conclude that $J_{NN}$ is AFM while the next-nearest-neighbor (NNN) interaction ($J_{NNN}$) is FM \cite{baker1966determination,brower1966electron}. These interactions form a triangle-based exchange network, as illustrated in Fig. \ref{fig:NuclearStructure}c,d.

Motivated by the proposed quasi-1D magnetic behavior for \NCns, which is unexpected based on the 3D connectivity of the lattice, and the lack of observed magnetic order, we have performed additional characterization on single crystals. Our low-temperature heat capacity measurements extend beyond the previous characterization and reveal two magnetic ordering transitions ($T_{N1}$ = 270 mK and $T_{N2}$ = 180 mK) below the previously reported broad feature ($T^{*}$ $\approx$ 450 mK). 

We probe the microscopic details of this staged magnetic ordering process through elastic neutron scattering measurements. Highly anisotropic magnetic diffuse scattering is observed at $T$ = 300 mK $> T_{N1}$ directly confirming the presence of strong quasi-1D short-range spin correlations and an effective dimensional reduction. Intriguingly, similar diffuse scattering persists in the intermediate regime $T_{N2} < T < T_{N1}$ where sharp antiferromagnetic Bragg peaks appear, indicating the coexistence of quasi-1D short-range order and 3D long-range order. In the low-temperature regime, $T$ = 45 mK $< T_{N2}$, full 3D long-range order is obtained with no change in magnetic propagation vector. We propose that the triangular-based exchange network in \NC leads to frustration of the exchange interactions that together with the presence of significant exchange anisotropy significantly impacts the magnetic phase behavior.

\section{Methods \label{methods}}
The bulk characterization measurements (magnetic susceptibility, magnetization and heat capacity) were performed on single crystals grown by slow cooling from a melt. \NC powder (Alfa Aesar, 99.99\%, ultra dry) was flame sealed as-received in a fused silica tube under vacuum. The tube was then hung in a vertical tube furnace and heated to 810 \textdegree C for 12 h before cooling at 0.4 \textdegree C/h to 700 \textdegree C for crystal growth, followed by cooling to room temperature at  60 \textdegree C/h. 

The elastic neutron scattering measurements were performed on a single crystal grown using a two-zone Bridgman furnace. Prior to growth, an open fused silica tube containing 25 g of anhydrous \NC powder was dried overnight at 300 \textdegree C under dynamic vacuum. The tube was then flame sealed under vacuum and hung in the furnace. The two zones, centered $\approx$ 20 cm apart, were set to 770 \textdegree C and 720 \textdegree C, respectively, and the tube was lowered through at 0.6 mm/h. 

\NC is very sensitive to moisture and thus all subsequent handling was performed in a glovebox. The boules resulting from both single crystal growth methods were found to be multi-domain, with apparent individual domain sizes reaching up to 1 cm. The crystals were mostly transparent with a color that depends on the lighting (green under fluorescent light, purple under natural light). While some sources suggest that the color change is due to different states of hydration, that is not the case. Room-temperature single crystal X-ray diffraction (SCXRD) data were collected on small pieces removed from the main boule using a Bruker D8 Advance Quest diffractometer with a graphite monochromator and Mo K$\alpha$ radiation ($\lambda$ = 0.71073 \r{A}).

Isolation of larger, individual, single-domain crystals proved challenging and the natural faceting that occurs upon breaking was not necessarily straightforward to identify.  We utilized a powder x-ray diffractometer (PANalytical X'pert Pro, Cu K$\alpha_1$ radiation, $\lambda$ = 1.5406 \r{A}) with an air-free holder to determine that the flat facets are typically built of ($1$ 0 0) planes; the presence of multiple facets on one crystal thus allows for identification of the [0 0 $1$] lattice direction by observation of an edge between two ($1$ 0 0) planes. On this basis, a single crystal (mass $\approx$ 185 mg, dimensions $\approx$ 2 mm $\times$ 2 mm $\times$ 5 mm) was selected for the elastic neutron scattering measurement during which the orientation was verified. A subsequent measurement of the magnetization of this well-oriented crystal with the magnetic field, $H$, applied along the [0 0 $1$] was used as a reference to verify orientation-dependent magnetic measurements on smaller crystals. 

For the bulk characterization measurements, the crystal was mounted in the glovebox as appropriate for the measurement and coated in grease to protect it from the atmosphere.  Magnetization and magnetic susceptibility measurements were performed on a crystal (mass = 15.5 mg) using the iHe3 insert in a Quantum Design MPMS3, with data collected upon warming after cooling in an applied field; isothermal magnetization data were collected upon demagnetization of the sample. Additional magnetic susceptibility data (not shown) were collected on powder sample (mass = 63 mg). The ac susceptibility (mass = 10.4 mg) and heat capacity (mass = 2.3 mg) were measured on a single crystal in a Quantum Design Dynacool using a dilution refrigerator (DR) insert. The magnitude of the ac susceptibility data did not match the dc approximation, $M$/$H$, which appears to be an  artifact of the Quantum Design AC-DR measurement system. As a result, we consider this measurement to be a qualitative probe for the occurrence of phase transitions and plot the data normalized to its value at \textit{T} = 1 K. For the $H \perp$ [0 0 $1$] magnetization and magnetic susceptibility measurements, the direction of $H$ in the \textit{ab}-plane was not controlled.

We utilized ChatGPT-5.6 Sol to construct a model for fitting the magnetization data. The model was based on the formalism by Bonner and Fisher for quantum 1D chains but modified to include $g$-factor anisotropy \cite{bonner1964linear}.  Further details are provided in the Supplemental Materials \cite{SI}.

Single-crystal elastic neutron scattering data were collected using the WAND$^{2}$ diffractometer at the High Flux Isotope Reactor (HFIR) using $\lambda$ = 1.486 \r{A} neutrons produced from the ($1$ $1$ $3$) Bragg peak of a Ge monochromater \cite{frontzek2018wand2}. The sample was mounted on a copper plate using thin copper foil and wire and oriented in the ($H$ $\bar{H}$ $L$) scattering plane. The assembly was then inserted inside a cylindrical copper sample vessel which was sealed using indium wire. All sample mounting was done inside a helium filled glovebox to avoid hydration and provide exchange gas ($P \approx $ 1 bar) within the sample vessel. The sample vessel was then mounted inside a dilution refrigerator capable of obtaining a base temperature of $\approx 45$ mK. 

Reciprocal space maps were obtained by rotating the sample 180\textdegree $ $ around the vertical axis ($\phi$) in steps of 0.1\textdegree. A temperature dependent order parameter was obtained on warming by stabilizing at each temperature (10 min at $T\leq200$ mK and 5 min at $T>$ 200 mK) before performing a limited $\phi$ rotation across the desired Bragg peak with the same step size as the reciprocal space maps. Peak integration for refinement of the nuclear and magnetic structures was done by summing the intensity within a region-of-interest in detector space for each observed peak as a function of $\phi$. The resulting intensity versus $\phi$ curve was then fit with a Gaussian plus constant background to extract the integrated intensity. A Lorentz factor correction was applied to the data. The peak intensities were also corrected for absorption with WinGX \cite{farrugia2012wingx} using a model that accounts for the sample shape. 

Rietveld refinements of the corrected elastic neutron scattering intensities were conducted with FullProf \cite{rodriguez2025} using lattice parameters determined from refinement of the UB matrix. A total of 61 integrated nuclear Bragg peaks and 44 integrated magnetic Bragg peaks were used in the refinements. The nuclear and magnetic peaks were refined separately, with the refined values of the scale factor, isotropic extinction factor, atomic positions, and atomic displacement parameters ($B_{iso}$) from the nuclear refinements fixed during the magnetic refinements. The nuclear structure was refined using different space groups, as discussed later. Refinement of the Nd $B_{iso}$ value was generally unstable, resulting in negative values or very large error bars, so this was fixed to a small, physically reasonable value (0.01 \r{A}$^{2}$). The Cl $B_{iso}$ value was more stable in the higher symmetry space groups, but not in the lower symmetry P3 space group, where it was subsequently fixed based on the higher symmetry space group refinements. In any case, refinement of the $B_{iso}$ values did not have a large impact on the goodness-of-fit or the resulting values of the other parameters. The magnetic structure refinements were conducted using the magnetic space group (MSG) approach. The associated symmetry analysis was conducted using the tool k-Subgroupsmag from the Bilbao Crystallographic Server \cite{perez2015symmetry}.

Inelastic neutron scattering measurements were performed using 4.1 g of the as-received \NC powder inside an Al sample vessel sealed with In wire under $\approx$ 1 bar of He exchange gas and loaded into a bottom loading closed cycle refrigerator sample changer with base temperature of 5 K \cite{Stone2025}. Measurements were conducted using the direct geometry spectrometer SEQUOIA at the Spallation Neutron Source (SNS) \cite{Stone2014}. Data were collected using a range of incident neutron energies ($E_i$) with the high-resolution Fermi chopper. For $E_{i}$ = 3, 6, 10, 25, 45, and 60 meV the Fermi chopper frequencies were 120, 120, 180, 240, 360, and 420 Hz, respectively. The measurement durations were defined by the amount of time for accumulation of a set number of Coulombs of charge on the spallation target and were 4.8, 22.2, 9.6, 14.4, 7.2, and 7.2 C, respectively (4.8 C of charge corresponded to approximately 58 minutes of data collection at the time of the measurement). Measurements were also made with $E_i$ = 500 mev using the high-flux Fermi chopper at a frequency of 480 Hz and a duration corresponding to 9.6 C. Measurements were performed with an empty sample vessel to provide an estimation of the background at select temperatures and incident energies. These measurements were performed with half the counting statistics as the sample measurements. To aid in identifying phonon contributions to the data, the expected phonon INS spectrum was calculated using the program INSPIRED \cite{han2024inspired,oclimax,phonopy1,phonopy2}.  

\section{Results and Analysis \label{results}}

\subsection{Bulk characterization}

 \begin{figure}
 \includegraphics[width = 8.4 cm]{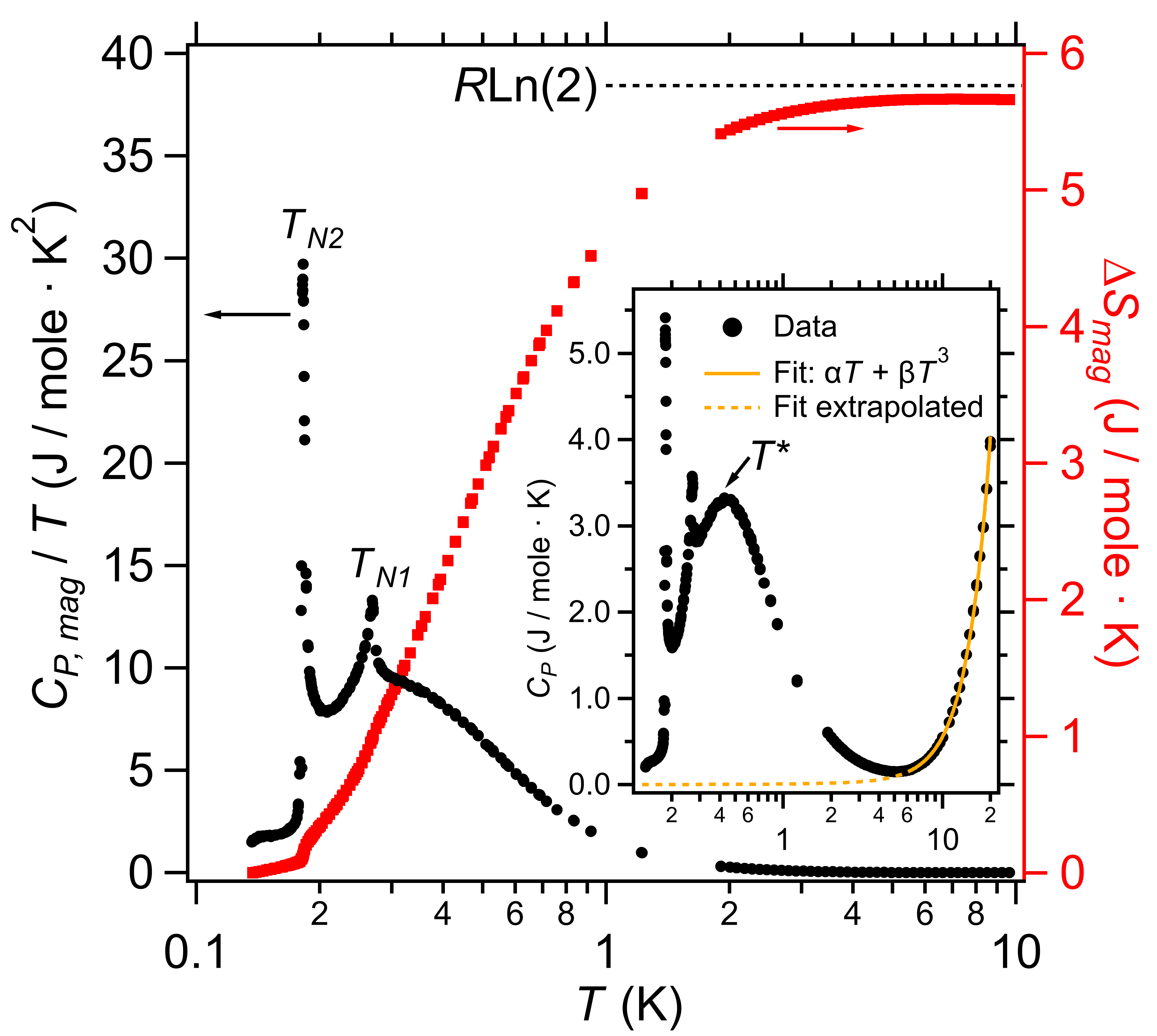}
 \caption{\label{fig:HeatCapacity} Magnetic contribution to the low-temperature heat capacity plotted as $C_{P,mag} / T$ (black circles, left axis) and the resulting integrated magnetic entropy release $\Delta S_{mag}$ (red squares, right axis). The dashed black line shows the position of $\Delta S_{mag}$ = $R$Ln(2). \textit{Inset:} Plot of the total heat capacity signal, $C_{P}$, highlighting the broad feature centered at $T^{*}$. The solid orange curve shows the fit  (described in the main text) used to estimate the non-magnetic contribution and the dashed orange curve is the extrapolation of the fit to the lowest $T$.}
 \end{figure}

The magnetic contribution to the low-temperature heat capacity, $C_{P,mag}$ $/$ $T$, is presented in the main panel of Fig. \ref{fig:HeatCapacity}, with the total heat capacity ($C_{P}$) in the inset. $C_{P,mag}$ was estimated by subtracting a non-magnetic contribution obtained by fitting the total heat capacity data across the range $T$ = 6 K -- 20 K to the expression $\alpha T + \beta T^{3}$. Sharp peaks are observed at $T_{N1}$ = 270 mK and $T_{N2}$ = 180 mK. No significant thermal hysteresis was observed for either feature. These peaks present upon cooling below the previously reported broad maximum centered at $T^{*} = 450 $ mK, the tail of which reaches nearly $T$ = 5 K \cite{colwell1969low,colwell1967low,eisenstein1965low}. 

The magnetic entropy release, $\Delta S_{m}(T) = \ \int_{0}^{T}{\frac{C_{p,mag}}{T'}dT'}$, plateaus to a value very close to \textit{R}Ln(2) by $T \approx$ 4.5 K and a large portion of the entropy release occurs above $T_{N1}$.  The net entropy change observed is consistent with the expected Kramer's doublet for Nd$^{3+}$. We note that estimating the non-magnetic contribution using fits to the full Debye model across a larger temperature range ($T$ = 6 K -- $T_{max}$, for various $T_{max}$) to isolate $C_{P,mag}$ gave essentially the same behavior for $\Delta S_{m}(T)$ and the result was independent of the precise fitted Debye temperature. 

 \begin{figure}
 \includegraphics[width = 8.4 cm]{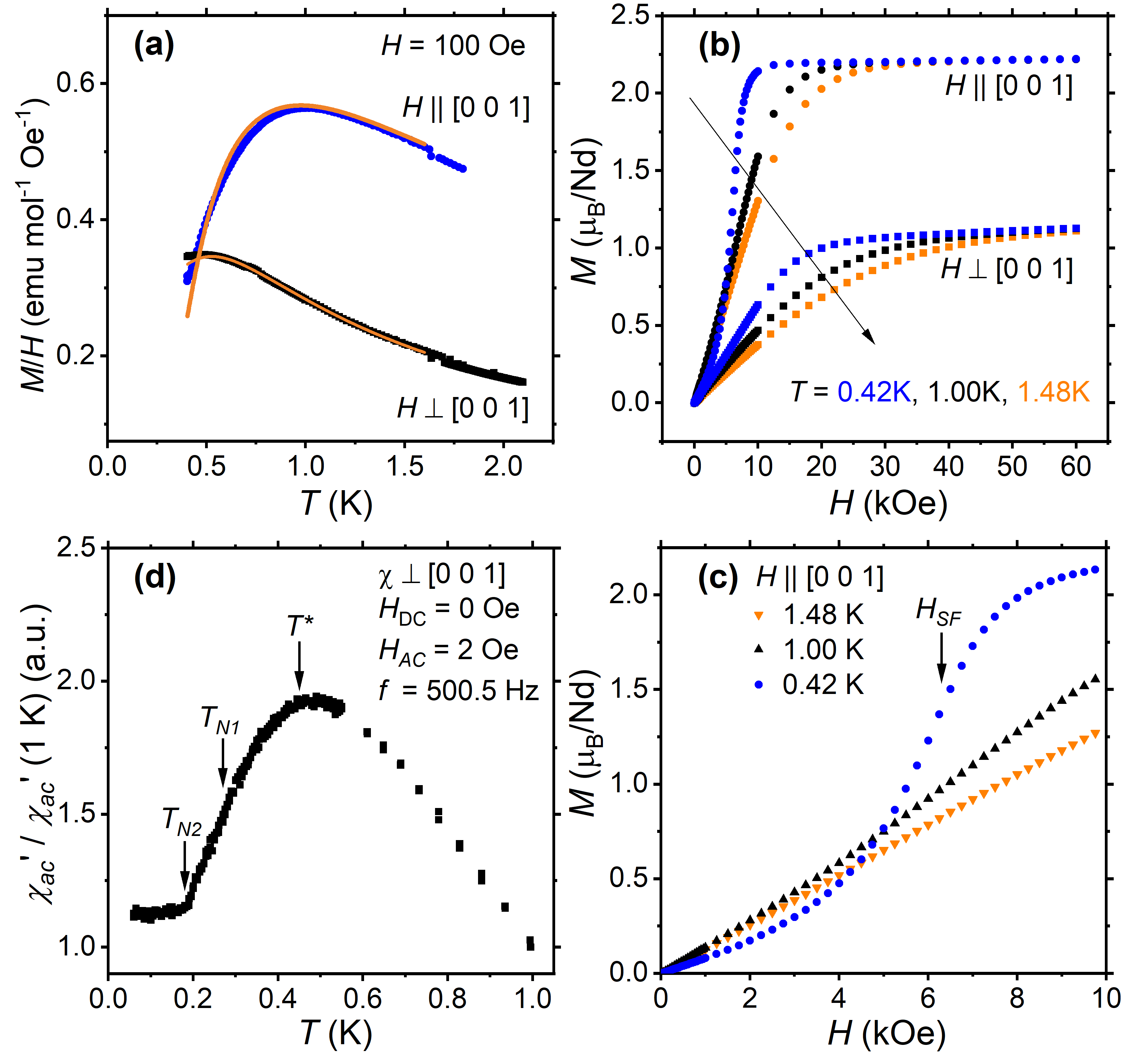}
 \caption{\label{fig:BulkMagnetic} Bulk magnetic characterization data collected on a single crystal with field parallel or perpendicular to the [0 0 $1$] lattice direction. (a) Temperature dependence of the dc magnetic susceptibility, \textit{M / H}. The solid orange lines are fits using the 1D XXZ model, as described in the main text. (b) Field dependence of the magnetization, $M$, at various temperatures. The arrow indicates the direction of increasing \textit{T} for both sets of measurements. (c) Low-field region of the $H$ $||$ [0 0 $1$] magnetization data. The arrow labels the position of the spin flop transition, $H_{SF}$, determined from the peak in the derivative with respect to field. (d) Temperature dependence of the ac magnetic susceptibility, $\chi'_{ac}$. The black arrows show the temperatures of the features observed in the heat capacity data. No feature corresponding to $T_{N1}$ is observed in $\chi'_{ac}$} 
 \end{figure}
 
Measurements of the dc magnetic susceptibility ($M/H$) down to $T$ = 0.42 K (Fig. \ref{fig:BulkMagnetic}a) reveal anisotropy between field applied in the $ab$-plane ($H \perp$ [0 0 $1$]) and along the $c$-axis ($H$ $||$ [0 0 $1$]). This magnetic anisotropy is manifested in two aspects. First, there is a variation in the magnitude of $M/H$, with a larger magnetization observed for $H$ $||$ [0 0 $1$] than for $H \perp$ [0 0 $1$]. This aspect is due to the known $g$-factor anisotropy caused by the nature of the crystal electric field. Second, the temperature at which a broad maximum occurs upon cooling varies significantly with orientation, with the maximum at a higher temperature for $H$ $||$ [0 0 $1$] than for $H \perp$ [0 0 $1$]. 

These results are qualitatively consistent with prior reports \cite{colwell1969low, eisenstein1965low}, which discussed this behavior as being driven by the emergence of quasi-1D correlations in a strongly anisotropic spin system. The existence of a broad maxima in $M/H$ is typically associated with short-range correlations developing in the paramagnetic state. In the case of \NCns, the presence of quasi-1D short-range order is directly confirmed by our elastic neutron scattering data, presented later. Accordingly, we analyzed the $M/H$ data using a 1D XXZ anisotropic exchange model for a chain of $S = \frac{1}{2}$ moments with only nearest-neighbor intrachain exchange (i.e. $J_{NN}$, with $J_{NN} < 0$ corresponding to AFM exchange). Specifically, we utilized the Bonner-Fisher formalism \cite{bonner1964linear} for 1D spin chains with XXZ exchange anisotropy. The Bonner-Fisher formalism was originally developed for a single, isotropic $g$-factor. To take into consideration the $g$-factor anisotropy present in \NC we modified the  Bonner-Fisher formalism by expanding the Zeeman term to include both $g_{||}$ and $g_{\perp}$. In the Bonner-Fisher formalism, the XXZ exchange anisotropy is represented by the variable $\gamma$, with  $\gamma$ = 0 being the Ising limit and $\gamma$ = 1 being the isotropic Heisenberg limit. The model utilizes exact diagonalization of the Hamiltonian of a ring of spins, and our convergence tests were consistent with those of Bonner and Fisher where a ring of 10 spins was found to be sufficient to make finite size effects negligible in the region of the broad maximum in the $M/H$ data. The fitting results for the range $T$ = 0.4 K -- 1.6 K (solid orange curves in Fig. \ref{fig:BulkMagnetic}a) demonstrate that the magnetic anisotropy is well described by this model in this temperature range. 

Our fitting produced $J_{NN}$ = -0.962(2)K and $\gamma$ = 0.152(2), confirming that \NC possesses exchange anisotropy close to the Ising limit as discussed previously \cite{colwell1969low, eisenstein1965low}. The best-fit parameters were also used to simulate the magnetic heat capacity of such a 1D chain, which is shown in the Supplemental Material \cite{SI}. The simulation gives a broad maximum centered very close to $T^{*}$, as observed in the heat capacity data (Fig. \ref{fig:HeatCapacity}, inset), albeit with a larger-than-observed peak value since the model does not capture the long-range order which occurs at lower temperature. We note that fitting the heat capacity data with the analytical result for the AFM Ising chain ($C_{P,Ising} = R(\frac{J_{NN}}{2k_{B}T})^{2}sech^{2}(\frac{J_{NN}}{2k_{B}T})$, where $R$ is the gas constant and $k_{B}$ is the Boltzmann constant) yields a similar result ($J_{NN}$ = -1.01(1) K) providing further indication that \NC is close to the Ising limit of exchange anisotropy.

A Curie-Weiss fit of dc magnetization data collected on a powdered single crystal between $T$ = 4 K -- 50 K (not shown) gave $\Theta_{C-W}$ = -1.37(3) K and $\mu_{eff}$ = 3.55(1) $\mu_{B}/$Nd. While this fitting range is sufficiently low to avoid modification of $\Theta_{C-W}$ due to the first excited CEF state, $\Theta_{C-W}$ is also influenced by the $g$-factor anisotropy of the ground state CEF doublet \cite{johnston2017influence} and the strong spin-orbit coupling \cite{li2021modified}, which complicates interpretation. The $\mu_{eff}$ value is consistent with the expectation for isolated Nd$^{3+}$ ($J$ = $\frac{9}{2}$, $g_{J} \approx$ 0.73, $\mu_{C-W}$ $\approx$ 3.62 $\mu_{B}$).

Magnetization versus field measurements (Fig. \ref{fig:BulkMagnetic}b,c) confirm that the easy axis of magnetization is along the $c$-axis. At $T$ = 0.42 K the ratio of the magnetization at $H$ = 60 kOe for the field parallel and perpendicular to the [0 0 $1$] ($M_{||}$ = 2.22 $\mu_{B}/$Nd and $M_{\perp}$ = 1.13 $\mu_{B}/$Nd, respectively) is $M_{||}$ $/$ $M_{\perp} \approx$ 1.97, similar to the ratio $g_{||}/g_{\perp}$ = 2.267(1) \cite{hutchison1958paramagnetic}. As emphasized in Fig. \ref{fig:BulkMagnetic}c, for $H$ $||$ [0 0 1] only, an inflection in the magnetization data appears at $H <$ 10 kOe for $T$ = 0.42 K (close to $T^{*} \approx$ 450 mK) prior to the beginning of saturation. This feature is consistent with a spin flop transition at $H_{SF} \approx$ 6.3 kOe, extracted using the position of the peak in the derivative with respect to field. 

Two clear features are visible in the ac susceptibility data, $\chi_{ac}'$ (Fig. \ref{fig:BulkMagnetic}d): a broad maximum close to $T^{*}$ and a kink at $T_{N2}$. A feature corresponding to the $T_{N1}$ peak observed in the heat capacity data is not observed in $\chi_{ac}'$. It may be that $T_{N1}$ manifests more subtly in $\chi_{ac}'$ and is hidden by the $T^{*}$ feature. Finally, we note that we did not measure $\chi_{ac}'$ as a function of applied frequency to inspect for glassy spin dynamics associated with the the broad $T^{*}$ feature observed. However, given the other available data consistent with quasi-1D behavior, we do not believe this feature is associated the with the formation of a spin glass phase.

\subsection{Inelastic neutron scattering}

\begin{figure*}
 \includegraphics[width = 17.4 cm]{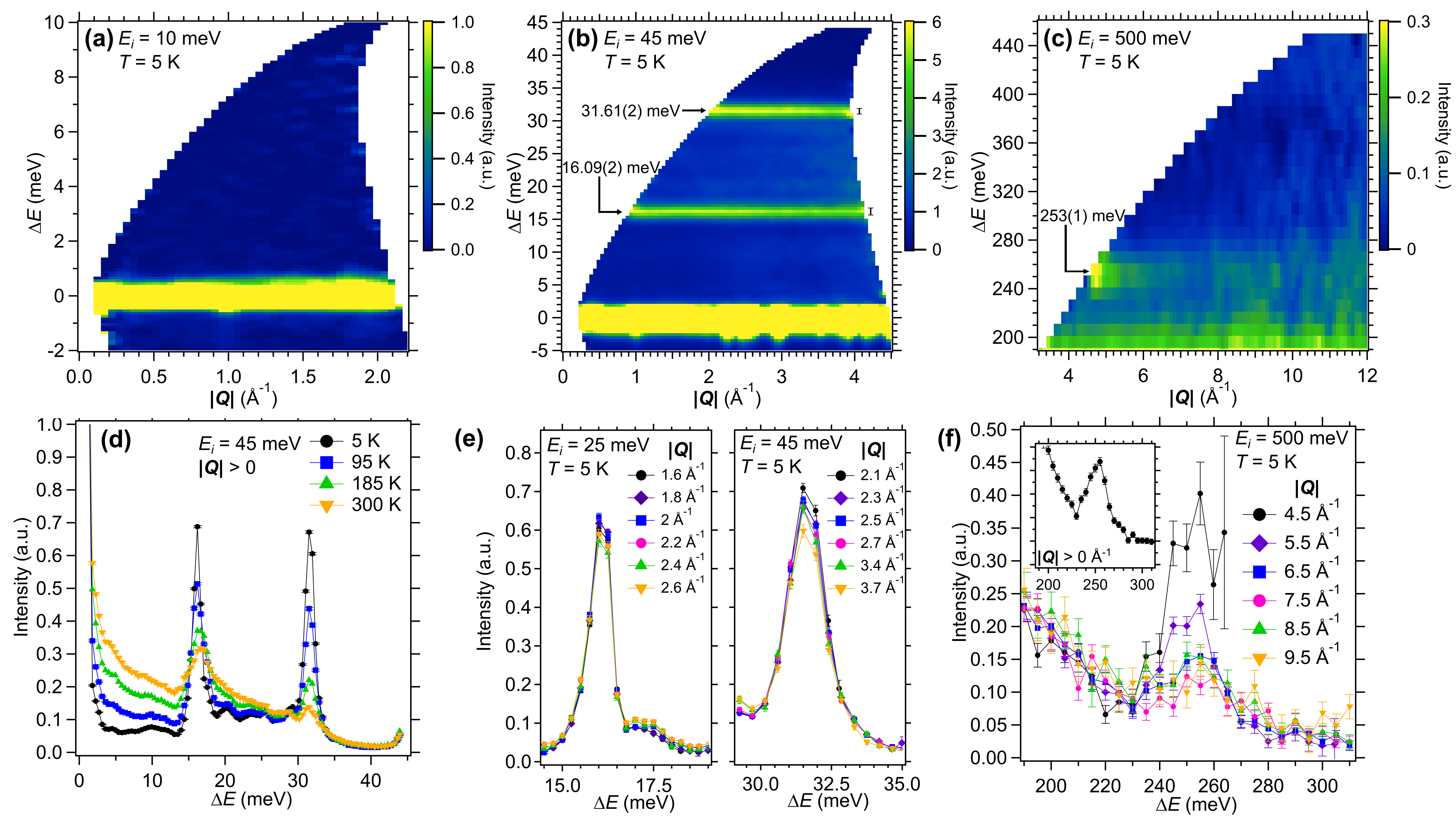}
\caption{\label{fig:INS} Inelastic neutron scattering data collected on a powder sample of \NC at various incident energies. The background from an empty can measurement has been subtracted in all plots. $\Delta E$ versus $|\vec{Q}|$ slices of the data are shown for (a) $E_{i}$ = 10 meV, (b) $E_{i}$ = 45 meV (intramultiplet CEF mode positions labeled by black arrows), and (c) $E_{i}$ = 500 meV (intermultiplet CEF mode labeled by black arrow). (d) Line cut along $\Delta E$ for $E_{i}$ = 45 meV with the data integrated over all $|\vec{Q}|$ values showing peaks corresponding to the two modes seen in (b). (e) Line cuts along $\Delta E$ at various $|\vec{Q}|$ for $E_{i}$ = 25 meV (left) and $E_{i}$ = 45 meV (right). The $|\vec{Q}|$ integration range is $\pm$ 0.1 \r{A}$^{-1}$ about each value in the legend. (f) Line cuts along $\Delta E$ at various $|\vec{Q}|$ for $E_{i}$ = 500 meV. The $|\vec{Q}|$ integration range is $\pm$ 0.5 \r{A}$^{-1}$ about each value in the legend. \textit{Inset}: Line cut along $\Delta E$ with the data integrated over all $|\vec{Q}|$.}
\end{figure*}

The INS data are consistent with the previously established well-isolated CEF doublet ground state \cite{carlson1961state,eisenstein1963spectrum}. Two dispersionless modes are observed centered at $\Delta E$ = 16.09(2) meV and 31.61(2) meV (Fig. \ref{fig:INS}a,b). Their magnetic origin is indicated by their temperature dependence, with both peaks decreasing in intensity as the temperature is increased (Fig. \ref{fig:INS}d). For the higher energy mode a clear decrease in intensity as a function of $|\vec{Q}|$ is also observed, as expected for a CEF mode. Calculations of the phonon spectra with the INSPIRED software package (see the Supplemental Material) indicate that a phonon overlaps with the lower energy CEF mode, causing the observed non-monotonic dependence of the intensity on $|\vec{Q}|$ (Fig. \ref{fig:INS}e, left). 

 The previous optical measurements established a CEF level scheme composed of five doublets consistent with the Kramers nature of the Nd$^{3+}$ ion with $J$ = $\frac{9}{2}$. The excited states are separated into two closely spaced pairs at $\Delta E$ = 14.3, 15.3, 30.3, and 30.9 mev \cite{carlson1961state,eisenstein1963spectrum}. The energy resolution of our measurement is 0.4 meV at $\Delta E$ = 16.1 meV for $E_{i}$ = 25 meV and 0.8 meV for $E_{i}$ = 45 meV at $\Delta E$ = 31.5 meV. Therefore, the observation of two CEF modes in the INS data is consistent with this level scheme, as we would not necessarily expect to be able to resolve the two modes of each pair separately. We do not attempt to use this data to analyze the ground state wave function because the low Nd site symmetry and our observation of only two modes means the CEF Hamiltonian is particularly underconstrained. In addition to these intramultiplet CEF modes, we also observe a higher energy mode at $\Delta E$ = 253(1) meV using $E_{i}$ = 500 meV (Fig. \ref{fig:INS}f). We identify this as the first intermultiplet transition ($J$ = $\frac{9}{2}$ $\rightarrow$ $\frac{11}{2}$) based on the observed decrease in intensity with $|\vec{Q}|$ and its consistency with the previous optical measurements \cite{carlson1961state,eisenstein1963spectrum}.

\subsection{Elastic neutron scattering}

\subsubsection{Nuclear scattering}

In this section we address the nuclear contribution to the elastic neutron scattering data. Reciprocal space maps are shown in Fig. \ref{fig:NeutronDiffraction}a -- d. We show only the low-$|\vec{Q}|$ region of this data as it captures the salient features of the nuclear contribution and is most relevant to the evolution of the magnetism, which we address in the next section. 

At $T$ = 10 K, in addition to the expected nuclear Bragg peaks for the room temperature space group, structure factor forbidden peaks are observed at positions with $H=K=0$ and $L$ = an odd integer (see orange circle in Fig. \ref{fig:NeutronDiffraction}a). These peaks persist unchanged across the full measured range, $T$ = 10 K -- 45 mK. The observation of weak, $T$-independent Bragg peaks in this temperature range which are forbidden by the room temperature space group is consistent with a structural distortion occurring at $T > $ 10 K. Parasitic $\lambda/2$ scattering can be ruled out because such scattering is absent for the (1 1 3) reflection of the Ge monochromator \cite{frontzek2018wand2}. An origin in a high density of coherent stacking faults is unlikely based on the connectivity of the structure. We also see no evidence of twinning which, due to the hexagonal symmetry, would most likely consist of domains rotated about the $c$-axis. Finally, given that the peaks occur only at well-defined, commensurate positions, scattering from an additional, misoriented grain in the sample is also unlikely. 

We  used the ISODISTORT package \cite{campbell2006isodisplace,stokesIsodisplace} to select the maximal subgroups of the room temperature structure (P6$_{3}$/m) which are consistent with the observed forbidden peaks. Consistent subgroups are those that can be reached via a $\vec{k}$ = 0 type lattice distortion and which allow intensity at the observed positions. This analysis indicates three compatible maximal subgroups: P$\bar{6}$ (\#174), P$\bar{3}$ (\#147), and Pm (\#6). P$\bar{6}$ and Pm only allow the forbidden Bragg peaks via the occurrence of a vacancy ordering-type distortion across the split (relative to the $T$ = 300 K structure) Nd and Cl Wyckoff positions. However, a substantial vacancy concentration was not observed in refinement of the $T$ = 300 K SCXRD data (not shown) indicating that vacancy ordering is not a viable explanation for the forbidden peaks and ruling out P$\bar{6}$ and Pm as the lowered symmetry. In contrast, P$\bar{3}$ allows intensity at the forbidden peaks via a displacement of the Nd and/or Cl ions along \textit{z}, away from the fixed \textit{z} = $\frac{1}{4}$ position in P6$_{3}$/m, lowering the Nd site symmetry to $C_{3}$. However, refinement of the $T$ = 45 mK data in P$\bar{3}$ gives a similar $R_{f}$-factor as P6$_{3}$/m and the resulting Nd and Cl atomic positions are identical within error in both refinements. 

Descending further in symmetry from the P$\bar{3}$ maximal subgroup, the only potential lower symmetry space group other than P1 is P3 (\#143), which has the same $C_{3}$ site symmetry for Nd. This space group splits the Nd and Cl sites and refinements produce larger displacements of the Nd/Cl ions along the \textit{z} direction, and therefore larger intensity at the forbidden (0 0 $L$)-type peaks. However repeated trials reveal that the refined values of the Nd and Cl \textit{z}-positions are biased by the initial value and that the statistical errors are always very large (of order 1). Furthermore, fixing the \textit{z}-positions to the symmetry constrained P6$_{3}$/m values and refining only the free Cl positions gives a nearly identical $R_{f}$ value. This refinement procedure leads to two Nd sites with the same site symmetry but a maximum difference of $\approx$ 1\% bond length difference between corresponding Nd-Cl bonds.

Overall, while our diffraction data qualitatively indicates a lower structural symmetry at $T \leq$ 10 K relative to $T$ = 300 K, the actual distortion is evidently quite subtle. Within the resolution of our measurement we cannot discern between P$\bar{3}$ or P3, which are the two most likely options. The results of the refinements of the $T$ = 45 mK data are summarized in Table \ref{tab:NuclearRefinementResults}. In the Supplemental Materials \cite{SI}, we show $|F^{2}|_{calc.}$ versus $|F^{2}|_{obs.}$ for these refinements, as well as for refinements of the $T$ = 205 mK and 10 K data using the same space groups. The same general behavior was observed for the higher temperature data as for the $T$ = 45 mK data, with all space groups offering equal quality of fits. We note here that P3 has been previously proposed to described the low-temperature structure of PrCl$_{3}$ based on Cl-NMR measurements \cite{hessler1971low}, which may support its assignment here. 

 \begin{table*}[!htbp]
 \centering
\renewcommand{\arraystretch}{1.4} 
 \caption{\label{tab:NuclearRefinementResults} Parameters for \NC refined from the $T$ = 45 mK neutron diffraction data using different space groups. All space groups belong to the hexagonal lattice system, although P3 is in the trigonal crystal system. A total of 61 integrated nuclear Bragg peaks were used for the refinement and the lattice parameters used were: \textit{a} = \textit{b} = 7.414(6) \r{A} and \textit{c} = 4.223(5) \r{A}. Symmetry constrained atomic positions are given as fractions while free positions are decimal values. Refined values possess error bars representing one standard deviation from the statistical output of the refinement. Values in \textit{italics} are free parameters which were fixed in the final refinement as discussed in the main text. Note that P6$_{3}$/m and P$\bar{3}$ have only one crystallographically distinct Nd and Cl site per unit cell, while P3 has two independent sites; this is indicated by the ``$\ast$'' label for the former two space groups.}
\begin{tabular*}{17cm}{c | c c c | c c c | c c c | c c c | c c | c }
\toprule
 Space group & \multicolumn{3}{c|}{Nd(1)}                                                               & \multicolumn{3}{c|}{Nd(2)}                                                   & \multicolumn{3}{c|}{Cl(1)}                             &  \multicolumn{3}{c|}{Cl(2)}               & \multicolumn{2}{c|}{$B_{iso}$ (\r{A}$^2$)} & {$R_{f}$-factor}  \\
             & \textit{x}                 & \textit{y}                  & \textit{z}                  & \textit{x}                  & \textit{y}                & \textit{z}          & \textit{x}     & \textit{y}    & \textit{z}             & \textit{x}  & \textit{y}  & \textit{z}    & Nd                & Cl                      &  \\ \hline
 P6$_{3}$/m, \#176 &  $\frac{1}{3}$             & $\frac{2}{3}$                & $\frac{1}{4}$               & $\ast$                      & $\ast$                    & $\ast$              & 0.3868(6)       & 0.3007(6)    & $\frac{1}{4}$          & $\ast$      &   $\ast$    & $\ast$        & \textit{0.01}\hspace{4pt}     & 0.13(7)             & 5.86\\ 
 P$\overline{3}$, \#147 &  $\frac{1}{3}$            & $\frac{2}{3}$                 & 0.250(1)                    & $\ast$                      & $\ast$                    & $\ast$              & 0.3868(8)       & 0.3009(8)     & 0.2487(5)             & $\ast$      &   $\ast$   & $\ast$         & \textit{0.01}\hspace{4pt}     & 0.15(8)               & 6.76 \\ 
 P3, \#143        &  $\frac{1}{3}$             & $\frac{2}{3}$                 & \textit{0.25}              & $\frac{2}{3}$                & $\frac{1}{3}$             & \textit{0.75}       & 0.381(3)        & 0.296(2)     & \textit{0.25}          & 0.608(2)      & 0.694(2)  & \textit{0.75} & \textit{0.01}\hspace{4pt}     & \textit{0.15}     & 6.82  \\
 \bottomrule
 \end{tabular*}
 \end{table*}

\subsubsection{Magnetic scattering}

\begin{figure*}
\includegraphics[width = 17.4 cm]{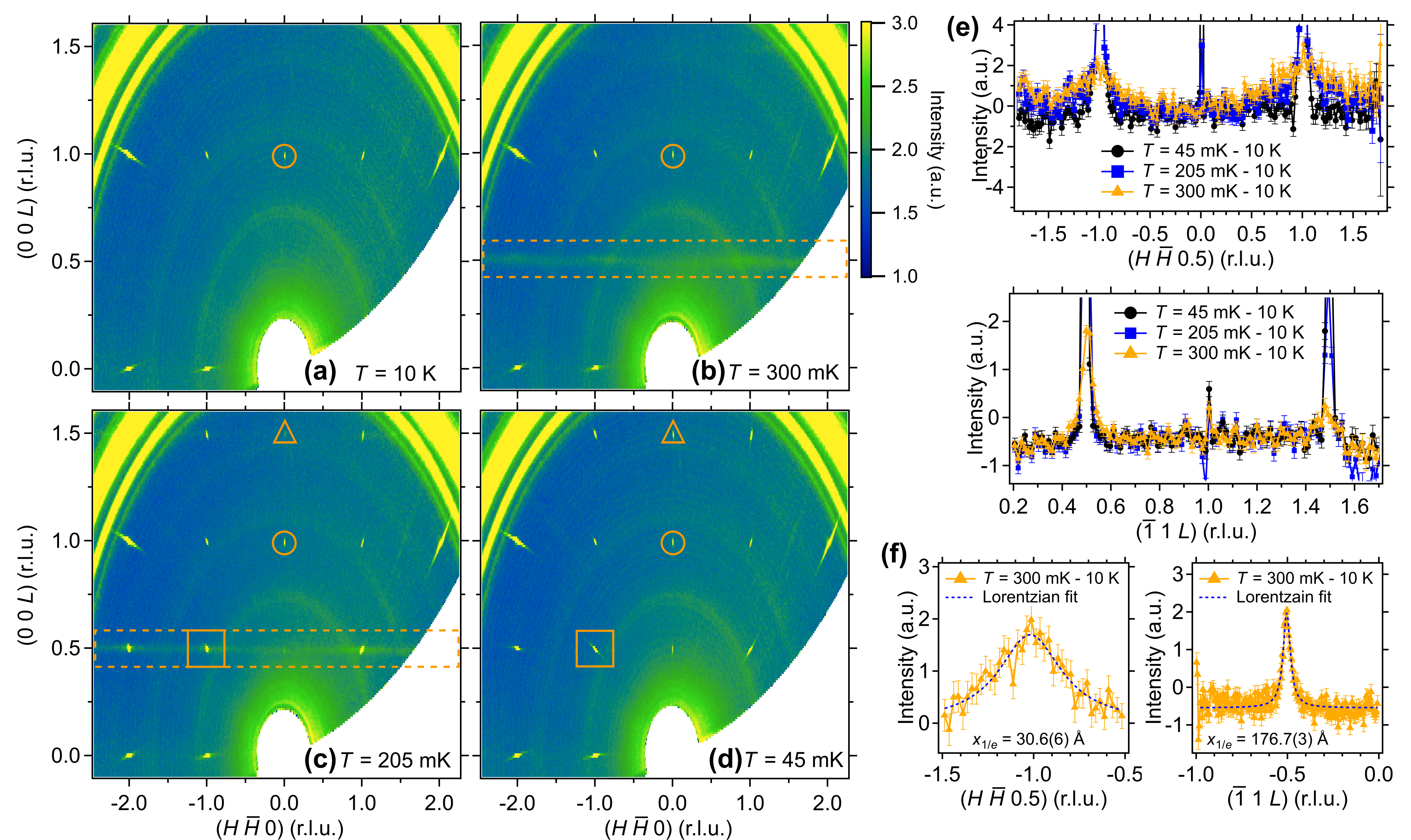}
\caption{\label{fig:NeutronDiffraction} Reciprocal space maps from single crystal neutron diffraction in the ($H$ $\bar{H}$ $L$) scattering plane collected at (a) 10 K $ >> T_{N1}$ (PM regime), (b) 300 mK $> T_{N1}$ (correlated PM regime), (c) $T_{N1} <$ 205 mK $< T_{N2}$ (intermediate regime), and (d) 45 mK $< T_{N2}$ (fully ordered regime). The orange solid and dashed lines highlight specific features discussed in the main text. In all cases the data was integrated across the full coverage of the out-of-plane ($H$ $H$ $0$) reciprocal lattice direction with the range ($H$ $H$ 0) = [-0.35, 0.35]. (e) Line cuts along ($H$ $\bar{H}$ 0.5) (upper) and ($\bar{1}$ $1$ $L$) (lower) at different temperatures following subtraction of the $T$ = 10 K data. For the cut parallel to ($H$ $\bar{H}$ 0) the integration ranges were (0 0 $L$) = [0.45, 0.55] and ($H$ $H$ 0) = [-0.05, 0.05]. For the cut parallel to the (0 0 $L$) direction the integration ranges were ($H$ $\bar{H}$ 0) = [-1.5, -0.5] and ($H$ $H$ 0) = [-0.05, 0.05]. (f) Fitting of the $T$ = 300 mK - 10 K line cut data shown in (e) at the $\vec{Q}$ = (1 $\bar{1}$ 0.5) position to a single Lorentzian (dashed blue line).}
\end{figure*}

We now turn to the magnetic contribution to the elastic neutron scattering data. As outlined below, the central features of the data define three main magnetic regimes: a correlated paramagnetic (PM) regime ($T >T_{N1}$), an intermediate regime ($T_{N2} < T < T_{N1}$) and a fully ordered regime ($T < T_{N2}$). At $T$ = 10 K, deep in the PM regime, no scattering is observed besides the nuclear Bragg peaks discussed previously (Fig. \ref{fig:NeutronDiffraction}a). We now consider each of the magnetic regimes in turn, from high to low-temperature. 

At $T$ = 300 mK $<T^{*}$ in the correlated PM regime, diffuse scattering emerges parallel to the ($H$ $\bar{H}$ 0) reciprocal lattice direction, centered at $L$ = half-integer positions. This is highlighted by the dashed orange box centered on $L$ = 0.5 in Fig. \ref{fig:NeutronDiffraction}b. An additional measurement at $T$ = 600 mK (not shown) reveals the same diffuse scattering, though with significantly weaker intensity. This indicates that the diffuses scattering develops in the neighborhood of $T^{*}$. The $T$-dependence of cuts along the $\vec{Q}$ = ($H$ $\bar{H}$ 0.5) and ($\bar{1}$ $1$ $L$) reciprocal lattice directions is shown in Figure \ref{fig:NeutronDiffraction}e. At $T$ = 300 mK the diffuse scattering is peaked at positions with integer  $H$ and  half-integer $L$, as expected from the reciprocal space map in Fig. \ref{fig:NeutronDiffraction}a. The cut along ($\bar{1}$ $1$ $L$) demonstrates that the intensity of the diffuse scattering decreases with increasing $L$, i.e. increasing $|\vec{Q}|$. The temperature and $|\vec{Q}|$-dependence together identify this as magnetic diffuse scattering. Note that because our neutron measurement is energy-integrated, we are unable to discern whether this diffuse scattering is static or dynamic in nature.

As demonstrated by the reciprocal space maps, the magnetic diffuse scattering is much narrower along the (0 0 $L$) direction than the ($H$ $\bar{H}$ 0). We extract the real space magnetic correlation lengths associated with this diffuse magnetic scattering by fitting cuts through the ($\bar{1}$ $1$ 0.5) position -- highlighted by the orange square in Fig. \ref{fig:NeutronDiffraction}c -- to a Lorentzian line-shape (Fig. \ref{fig:NeutronDiffraction}f). Incorporating the instrumental resolution via the Gaussian width of a nearby nuclear Bragg peak and a pseudo-Voigt line-shape does not meaningfully affect these results because the nuclear Bragg peaks are much narrower than the diffuse scattering. The correlation length is $x_{1/e}$, defined as the 1/$e$ decay length of the Fourier transform of the Lorentzian line-shape. From this analysis we obtain $x_{1/e}$ = 31(3) \r{A} along ($H$ $\bar{H}$ 0) and $x_{1/e}$ = 180(11) \r{A} along (0 0 $L$). Analogous slices and cuts to those shown here indicate that the diffuse signal along the orthogonal direction ($H$ $H$ 0) is similarly broad as in the ($H$ $\bar{H}$ 0) direction. Although the limited coverage in the ($H$ $H$ 0) direction precludes the same quantitative analysis of the correlation length, it is clear that the diffuse scattering presents in reciprocal space as a plane defined by the ($H$ $\bar{H}$ 0) and ($H$ $H$ 0) directions, with the normal direction to the plane parallel to (0 0 $L$). 

\begin{figure}
 \includegraphics[width = 8.4 cm]{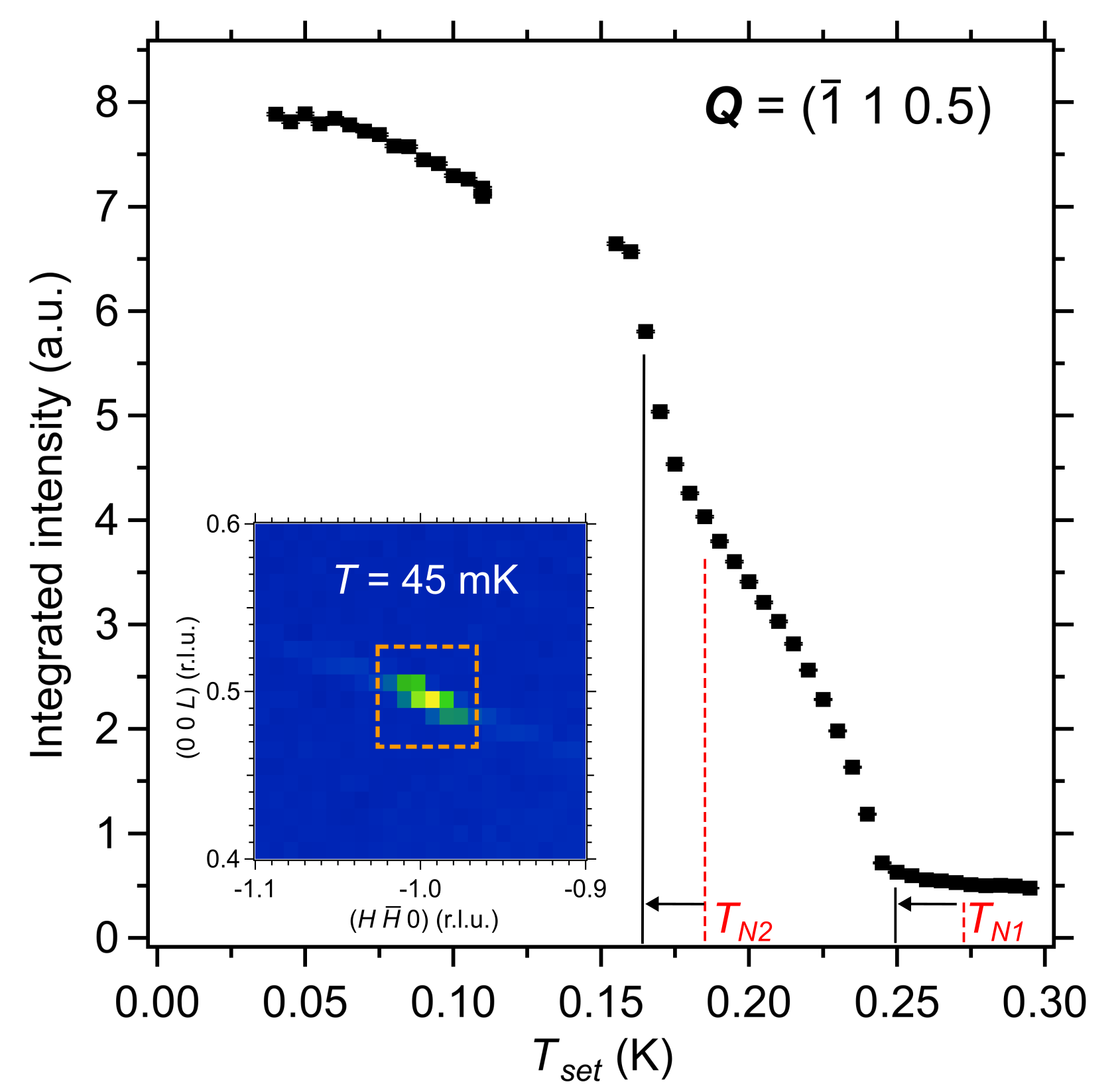}
 \caption{\label{fig:NeutronOrderParameter} Temperature dependence of the integrated intensity of the ($\bar{1}$ $1$ 0.5) magnetic Bragg peak. The dashed red lines labeled $T_{N1}$ and $T_{N2}$ show the position of the peaks in the heat capacity data. The solid black lines are shifted by the same amount to lower temperature, with the shift equal to the distance between the $T_{N1}$ temperature determined from heat capacity and the temperature where the Bragg peak intensity first increases sharply. \textit{Inset}: View of the Bragg peak in momentum space with the integration region indicated by the dashed orange box.}
 \end{figure}
 
The intermediate regime ($T_{N2} <$ $T$ = 205 mK $< T_{N1}$) is defined by the emergence of new Bragg peaks at ($H$ $\bar{H}$ $L$) positions with integer $H$ and half-integer $L$ (see e.g. Fig. \ref{fig:NeutronDiffraction}c, orange square). In Fig. \ref{fig:NeutronOrderParameter} we show the integrated intensity of the ($\bar{1}$ $1$ 0.5) position as a function of the nominal sample temperature ($T_{set}$). There is a clear increase in the intensity over the background near $T_{set}$ = 250 mK, marking the entrance into the intermediate-temperature regime. Up to a constant temperature offset this increase maps to $T_{N1}$ from the heat capacity data, as indicated by the black and red lines in Fig. \ref{fig:NeutronOrderParameter}. The offset in the neutron data reflects a small difference in the actual sample temperature from $T_{set}$ due to the distance of the sample from the dilution refrigerator mixing chamber. These new Bragg peaks with half-integer $L$ decrease in intensity with increased $|\vec{Q}|$; this, in addition to the observed $T$-dependence, confirms they stem from magnetic order. Their position in reciprocal space defines this magnetic order as antiferromagnetic (AFM) with a propagation vector of $\vec{k}$ = (0 0 $\frac{1}{2}$).

The reciprocal space map at $T$ = 205 mK demonstrates that the diffuse scattering persists in this regime, coexisting with the magnetic Bragg peaks. To determine the magnetic correlation length along the $c$-axis we analyzed a cut through the (-1.2 1.2 0.5) position with integration ranges ($H$ $\bar{H}$ 0) = [-1.3, -1.1] and ($H$ $H$ 0) = [-0.1, 0.1] in order to avoid the impact of the much stronger Bragg peak scattering. This cut (not shown) demonstrates that the diffuse scattering narrows along (0 0 $L$) upon cooling from the correlated PM regime to the intermediate regime. The same Lorentzian line-shape analysis yields an increased correlation length of 400(44) \r{A}. It is not possible to avoid the strong Bragg scattering along ($H$ $\bar{H}$ 0) so we do not repeat this analysis for that direction.

\begin{table}
  \centering
  \setlength{\tabcolsep}{10pt}   
  \renewcommand{\arraystretch}{1.2} 
    \caption{Correlation lengths (Lorentzian 1/$e$ decay length) from the analysis of the diffuse scattering at different temperatures and along different reciprocal space directions, as described in the main text. The entry ``\textit{N/A}'' indicates that no analysis was done for that temperature and reciprocal space direction.}
  \label{tab:DiffuseAnalysis}
  \begin{tabular}{c c | c c}
    \toprule
    \multicolumn{2}{c|}{$T$ (mK)} & \multicolumn{2}{c}{$x_{1/e}$ (\r{A})} \\
      & & ($H$ $\bar{H}$ 0) & (0 0 $L$) \\ \hline
    \multicolumn{2}{c|}{300} & 31(3) & 180(11) \\
    \multicolumn{2}{c|}{205} & \textit{N/A} & 400(44) \\
    \bottomrule
  \end{tabular}
\end{table}

Finally, in the fully ordered regime at $T$ = 45 mK (Fig. \ref{fig:NeutronDiffraction}) the magnetic diffuse scattering disappears entirely and only the same AFM Bragg peaks which marked the emergence of the intermediate regime remain. While no changes in the positions of the Bragg peaks are observed, there is an overall increase in intensity relative to $T$ = 205 mK. This is seen in the order parameter measurement (Fig. \ref{fig:NeutronOrderParameter}) as an additional increase across the range $T_{set} \approx$ 175 mK -- 160 mK. This feature maps to $T_{N2}$ in the same manner as the higher temperature feature maps to $T_{N1}$.

We analyzed the line-shape of the ($\bar{1}$ $1$ 0.5) magnetic Bragg peak at $T$ = 45 mK via Gaussian fits to the cuts shown in Fig. \ref{fig:NeutronDiffraction}e. This analysis yields full-width-at-half-max (FWHM) values of 0.0116(1) r.l.u. along (0 0 $L$) and 0.0215(3) r.l.u. along ($\bar{H}$ $H$ 0). An estimate of the instrumental resolution was obtained by fitting the nearby ($\bar{2}$ $2$ $1$) nuclear Bragg in the same manner, yielding FWHM's of 0.01174(2) r.l.u. along (0 0 $L$) and 0.02320(4) r.l.u. along ($H$ $\bar{H}$ 0). Therefore the ($\bar{1}$ $1$ 0.5) magnetic Bragg peak is resolution limited along both directions at $T$ = 45 mK. In conjunction with the saturation of the order parameter at this temperature, this analysis indicates that, within the resolution of our measurement, the long-range magnetic order is fully formed and a magnetic structure solution can be attempted.

In order to refine the magnetic structure via the MSG approach all of the magnetic subgroups of the nuclear (paramagnetic) space group which are consistent with the magnetic propagation vector, $\vec{k}$ = ($0$ $0$ $\frac{1}{2}$), should be considered. For \NCns, the maximal magnetic subgroups of each of the considered nuclear space groups require that the moments be purely along the $c$-axis. However, as we discuss later, the observed distribution of magnetic Bragg scattering in reciprocal space indicates that there is a moment component in the $ab$-plane as well as along the $c$-axis. Descending further in symmetry we found only one magnetic subgroup which allows a moment component in the $ab$-plane: P$_{\textrm{b}}$m (\# 6.22), which is a non-maximal subgroup of the nuclear space group P6$_{3}$/m. This magnetic space group enforces two Nd sublattices with independent moment magnitudes and directions: one sublattice is strictly parallel to the $c$-axis ($\vec{m}_{c}$) while the other is constrained to the $ab$-plane ($\vec{m}_{ab}$). Examining the $T$ = 205 mK data leads to the same magnetic space group selection.

Refinement of the $T$ = 45 mK data in the P$_{\textrm{b}}$m magnetic space group with $\vec{m}_{ab}$ along the [1 1 0] direction gave a magnetic structure with substantially different magnitudes for $\vec{m}_{c}$ and $\vec{m}_{ab}$. In the hexagonal coordinate system the magnetic moments on the two sublattices are: $\vec{m}_{c}$ = [0 0  3.1(1)] $\mu_{B}$ and $\vec{m}_{ab}$ = [0.60(5) 0.60(5) 0] $\mu_{B}$. From a fitting perspective this solution captures the data reasonably well ($R_{f}$ = 10.3). Refinement of the $T$ = 205 mK data with this model yielded a qualitatively similar magnetic structure solution. However, the goodness-of-fit was substantially worse, presumably due to the overall lower Bragg peak intensities and the presence of the diffuse scattering.

In the Appendix we provide further investigations into fitting the $T$ = 45 mK data in order to probe the robustness of this type of magnetic structure solution. Despite being robust from a fitting perspective, such a magnetic structure is hard to rationalize given the currently available experimental data from other probes -- as we address in the Discussion section -- and we do not consider it to be definitive. 

\section{Discussion \label{discussion}}
From the elastic neutron scattering and bulk characterization results it is clear that \NC goes through a staged magnetic ordering process defined by three regimes. To recap, these are the correlated PM regime ($T >T_{N1}$), the intermediate regime ($T_{N2} < T < T_{N1}$), and the fully ordered regime ($T < T_{N2}$). The observation of two peaks in the heat capacity data confirms that these regimes are separated by phase transitions at $T_{N1}$ and $T_{N2}$. Additionally, the presence of features in the neutron order parameter measurement (Fig. \ref{fig:NeutronOrderParameter}) which correspond to the peaks in the heat capacity data  identifies these phase transitions as magnetic in nature. A partial magnetic ordering occurs on entering the intermediate regime and then develops further in the fully ordered regime. We note that the temperature scale of these transitions is well-matched to a previous analysis of $^{35}$Cl nuclear quadrupole resonance data, which predicted long-range order in the range $T$ = 0.17 - 0.3 K \cite{colwell1969low}. As in the previous section, we will now discuss each regime in turn, proceeding from high to low-temperature.

The diffuse magnetic scattering in the correlated PM regime indicates the formation of short-range, magnetic correlations in real space. Since the diffuse scattering forms planes in reciprocal space perpendicular to the (0 0 $L$) reciprocal lattice direction (which present as lines in the reciprocal space maps), the associated real space short-range magnetic correlations are quasi-1D. The correlation length analysis confirms this explicitly, showing much larger lengths along the $c$-axis than in the $ab$-plane. A correlated PM state is consistent with the presence of a significant magnetic entropy release in this regime as the expected plateau to $R$Ln(2) only occurs at $T\approx$ 4 K, well above $T_{N1}$ and coinciding with the tail of the $T^{*}$ feature. Since \NC is confirmed to have a well-isolated doublet ground state, ruling out a Schottky origin, the $T^{*}$ feature is clearly tied to these quasi-1D magnetic correlations. The appearance of a corresponding $T^{*}$ feature in the $\chi_{ac}'$ data further corroborates the magnetic origin. Additionally, since the diffuse intensity is maximized at $L$ = half-integer positions, the associated magnetic correlations have the same AFM nature as the long-range order which sets in at lower temperature. Such quasi-1D AFM correlations are consistent with the successful fitting of the $M/H$ data in this temperature range using an AFM 1D model (Fig. \ref{fig:BulkMagnetic}a). 

These magnetic correlations are unlikely to be driven by an inherently quasi-1D exchange network given that the lattice has a clear 3D connectivity. As mentioned previously, ESR data indicates that the intrachain $J_{NN}$ is AFM (in agreement with our analysis of the magnetic susceptibility) while the interchain $J_{NNN}$ is FM \cite{baker1966determination,brower1966electron}. The coordination for $J_{NN}$ is along the $c$-axis, aligning with our observation of short-range AFM correlations along this direction. It is reasonable to assume that $|J_{NN}|$ $\approx 1$ K $> |J_{NNN}|$ since the bond angles of the associated Nd-Cl-Nd superexchange pathways are the same and the NN real space distance (4.2465(3) \r{A} from $T$ = 300 K SCXRD) is approximately 10\% shorter than the NNN real space distance (4.7719(1) \r{A}). Additionally, $J_{NN}$ is associated with face-sharing connections between Nd polyhedra while $J_{NNN}$ is associated with edge-sharing connections, which also suggests $|J_{NN}| > |J_{NNN}|$ \cite{kugel2015spin}. While it is beyond the scope of this paper to determine $|J_{NNN}/J_{NN}|$ precisely, the difference in magnitude is unlikely sufficient to generate the quasi-1D behavior we observe. For comparison, materials with structurally well-isolated spin chains typically have intrachain exchange interactions which are an order-of-magnitude larger than the interchain ones \cite{yasuda2005neel,ami1995magnetic,lancaster2006magnetic,satija1980neutron}. 

Instead, we ascribe this quasi-1D behavior to an effective dimensional reduction due to the presence of frustration. Examining the exchange interaction network shows that the offset between next-nearest-neighbor Nd ions means the AFM chains defined by $J_{NN}$ are bipartite. This is represented by the different colored Nd ions in Fig. \ref{fig:NuclearStructure}c, with the FM interchain $J_{NNN}$ always connecting chains of different color. The exchange network is  composed of isosceles triangles with one AFM $J_{NN}$ side and two FM $J_{NNN}$ sides (gray shaded region in Fig. \ref{fig:NuclearStructure}c). The dominance of $J_{NN}$ is effectively enhanced because the tendency toward AFM correlations within each chain partially frustrates the FM interchain $J_{NNN}$. In Fig. \ref{fig:NuclearStructure}d the $J_{NNN}$ exchange bonds are divided into two subsets (dashed and solid green lines); given the $\vec{k}$ = (0 0 $\frac{1}{2}$) AFM correlations, it is not possible to fully satisfy both subsets at the same time. This effectively decouples the AFM chains defined by $J_{NN}$, leading to quasi-1D magnetic behavior. Note that for AFM $J_{NN}$ this frustration mechanism is operative regardless of whether $J_{NNN}$ is FM (as for \NCns) or AFM. Such a situation is similar in spirit to the frustration in the planar anisotropic triangular lattices of NaMnO$_{2}$ \cite{dally2018amplitude} and Ca$_{3}$ReO$_{5}$Cl$_{2}$ \cite{hirai2019one} mentioned previously, although the overall geometry and details of the exchange interactions are different.

The coexistence of diffuse scattering and AFM Bragg peaks in the intermediate regime is likely also associated with frustration. Such a coexistence implies a partial ordering, where some portion of the Nd moments have obtained long-range order while another portion remains short-range ordered. Behavior of this type (sometimes referred to as ``partially disordered antiferromagnetism'') is known to occur in Ising models on the triangular lattice, as exemplified by CsCoCl$_{3}$, which possesses effective AFM spin chains running perpendicular to a planar triangular lattice   \cite{kudasov2009magnetic,mekata1977antiferro,mekata1978magnetic,miyashita1986magnetic,plumer1989landau,koseki2000monte, todoroki2004ordered}. In CsCoCl$_{3}$ there is an intermediate temperature phase wherein two out of the three chains per triangle order AFM while the third remains disordered and only develops an ordered moment at lower temperature. While the triangle-based exchange network in \NC is distinct from the planar triangular lattice, we conjecture that, given the clear evidence for Ising-like exchange anisotropy, the partially disordered magnetic state we observe is also stabilized by frustration. 

Addressing the nature of the long-range magnetic order, based on the propagation vector the naive expectation is a uniform Nd moment magnitude which forms AFM chains along the $c$-axis. Due to the relative shift between the chains, half of the NNN Nd ions in the neighboring chains would be parallel and half would be antiparallel. The distribution of the magnetic Bragg scattering in reciprocal space provides information about the expected moment direction. An important point is the presence of AFM Bragg peaks of the type ($0$ $0$ $L$), such as the ($0$ $0$ $1.5$) marked by an orange triangle in Fig. \ref{fig:NeutronDiffraction}c and d. The presence of this type of peak indicates that a finite component of the moment lies in the $ab$-plane since the magnetic elastic neutron scattering cross-section is only sensitive to the component of the moment perpendicular to $\vec{Q}$. However, comparing Bragg peaks with similar $|\vec{Q}|$ shows that those with a finite ($H$ $\bar{H}$ 0) component are more intense than those without. For example, the ($\bar{1}$ $1$ $1.5$) magnetic Bragg peak is much more intense than the ($0$ $0$ $1.5$). The similar values of $|\vec{Q}|$ (2.44 \r{A}$^{-1}$ and 2.23 \r{A}$^{-1}$, respectively) indicate that the intensity difference is due to the moment direction, not the magnetic form factor, and that that the total moment is predominately along the $c$-axis. Therefore, the expectation is that the Nd moments are canted and make a small angle with the $c$-axis. Such an ordering would be consistent with the $c$-axis easy  direction inferred from the ESR data \cite{hutchison1958paramagnetic} and our bulk magnetic characterization, particularly the observation of a spin flop transition with $H$ parallel to [0 0 $1$] (Fig. \ref{fig:BulkMagnetic}c). Since the magnetic propagation vector remains consistent and there is no dramatic redistribution of the magnetic Bragg peak intensity (beyond the overall increase at lower $T$), the nature of the long-range magnetic order is most likely largely unchanged between the intermediate and fully-ordered regimes. 

The magnetic structure refined at $T$ = 45 mK clearly deviates from this naive expectation. Rather than canted moments, it possesses two orthogonal magnetic sublattices (along the $c$-axis and in the $ab$-plane, respectively) with substantially different moment magnitudes. As we discuss below, this solution is not completely physically consistent and we do not consider it be definitive. From a symmetry perspective there is no issue with the distinct moment directions since, generically, the operators of a magnetic space group can break the equivalence of two sites which are the same at the crystallographic level. This allows the magnetic sublattices in the utilized magnetic space group to have orthogonal direction constraints. While distinct moment directions are somewhat unexpected in light of the observed uniform uniaxial anisotropy of the two sites, the pure $ab$-plane moment is not manifestly nonphysical since the anisotropy is not extremely strong ($g_{||}/g_{\perp} \approx 2.3$). Conceivably, such a difference in directions might be stabilized by the combination of frustration and exchange anisotropy.

In contrast, the distinct moment magnitudes are difficult to justify physically. In an insulating rare-earth magnet the moment magnitude is typically a single-ion property impacted by the nature of the local coordination environment and the associated CEF ground state and $g$-factor. Here, the refined value of $\vec{m}_{c}$ (3.1(1) $\mu_{B}$) is larger than the value allowed by the $g$-factor measured at $T$ = 4 K: $g_{||}J_{eff} = g_{||}/2 \approx$ 2 $\mu_{B}$. While $|\vec{m}_{ab}| < g_{\perp}J_{eff}$ could be reconciled by the existence of persistent fluctuations, $|\vec{m}|_{c} > g_{||}J_{eff}$ cannot be. Furthermore the refined moment sizes are inconsistent with the magnitude of the saturated moment from the low-temperature magnetization data (Fig. \ref{fig:BulkMagnetic}b); in contrast, the magnetization data is consistent with the reported $g$-factor anisotropy. For these reasons, we are unable to conclude that this refined magnetic structure represents the true magnetic ground state of \NCns. While our analysis suggests a qualitatively similar ordered magnetic structure is present in the intermediate regime, the aforementioned issues with this solution prevent this conclusion as well. We consider this aspect of the magnetism to be an open question that needs to be addressed through future studies. 

Having proposed that inherent frustration in the exchange network of \NC is relevant to the staged magnetic ordering process it is useful to compare our results to those from structurally similar rare-earth-based magnetic materials. The Sr$RE_{2}$O$_{4}$ family provides one such example, as the $RE$ ions form an offset chain structure similar to that of \NC \cite{fennell2014evidence,gauthier2017absence,hayes2011coexistence,petrenko2008low,quintero2012coexistence,wen2015disorder,young2012low,young2013highly}. However, in these materials the crystal structure possesses multiple $RE$ sites with significant differences in the symmetry of their local coordination environments and associated CEF ground states. This feature has been central to explanations of the magnetic phase behaviors across the family, which also include quasi-1D magnetism and the coexistence of long-range and short-range spin correlations. These similarities are surprising given that any distinction between Nd sites in \NC is evidently quite subtle. This suggests that these magnetic phase behaviors do not require multiple distinct rare-earth sites to occur, thus reinforcing the role of frustration. Another structurally similar material is the intermetallic DyNi$_{5}$Ge$_{3}$, which recent work has demonstrated possesses quasi-1D magnetism and an intermediate coexistence regime prior to full long-range order \cite{ge2022successive}. The observation of these common magnetic phase behaviors across chemically and electronically distinct materials also suggests they may be characteristic of this particular triangle-based exchange network. Theoretical studies and modeling dedicated to the $T$-dependent phase diagram of this exchange network are needed to address this conjecture.

There are also clear motivations for further experimental work. Additional low-temperature, high-resolution elastic neutron scattering studies providing a broader survey of reciprocal space would be useful to address the ambiguity in the nuclear structure and resolve the question of the magnetic ground state. A polarized elastic neutron scattering experiment may be needed to address the magnetic ground state, as well as to discern if there is any structural component to the observed transitions. It may also be informative to measure the $g$-factors on pure \NC crystals, rather than dilute (La$_{1-x}$Nd$_{x}$)Cl$_{3}$ crystals, as done previously \cite{hutchison1958paramagnetic}. The CEF experienced by Nd in the dilute samples is likely slightly different from that in pure \NCns, which could affect the $g$-factor. Measurement of the frequency dependence of $\chi_{ac}'$ could also be useful to check for any glassiness occurring across the different temperature regimes. Finally, single-crystal inelastic neutron scattering measurements would be valuable to discern the nature of the diffuse scattering (static versus dynamic), particularly in the intermediate regime.

\section{Conclusions \label{conclusions}}

Our elastic neutron scattering and bulk characterization data reveal the existence of multiple magnetic regimes in \NCns. In the correlated PM regime the system shows short-range, AFM, quasi-1D magnetic correlations, representing an effective dimensional reduction given the 3D lattice connectivity. The presence of significant magnetic correlations in this regime explains the broad feature at $T^{*}$ in the heat capacity and magnetic susceptibility, as well as the accompanying large magnetic entropy release. In the intermediate regime, magnetic Bragg peaks described by a $\vec{k}$ = (0 0 $\frac{1}{2}$) propagation vector emerge. However, the quasi-1D diffuse scattering persists in this regime, indicating only a partial magnetic ordering. We propose that these phases behaviors are linked to frustration of anisotropic exchange interactions present in this system. In the fully ordered regime, the disappearance of the diffuse scattering and increase in the AFM Bragg peak intensity indicates that complete long-ranged order is obtained with the same $\vec{k}$. The refined AFM magnetic structure which best captures the observed magnetic Bragg peak intensities appears nonphysical, and the true magnetic ground state remains an open question.

\section{Data Availability}

The data that support the findings of this article are publicly available \cite{data}. Further details are available upon reasonable request to the authors.

\begin{acknowledgments}
We thank Gabrielle Sala for assistance with the neutron scattering experiments and helpful discussion. We also acknowledge G\'abor Hal\'asz and Pyeongjae Park for useful discussions. This research was supported by the U.S. Department of Energy, Office of Science, Basic Energy Sciences, Materials Science and Engineering Division (experiment planning, data collection, data analysis, and manuscript preparation). This research used resources at the High Flux Isotope Reactor and Spallation Neutron Source, DOE Office of Science User Facilities operated by the Oak Ridge National Laboratory. The beam time was allocated to WAND$^2$ on proposal number IPTS-34499.1 and to SEQUOIA on proposal number IPTS-22774.1.
\end{acknowledgments}

\appendix
 \renewcommand{\thefigure}{A\arabic{figure}}
 \renewcommand{\thetable}{A\Roman{table}}
 \renewcommand{\thesection}{}
 \setcounter{figure}{0}
 \setcounter{table}{0}
\section{}

Given the difficulties with the magnetic structure solution presented above, here we provide further analysis of the $T$ = 45 mK elastic neutron scattering data. To examine whether the features of distinct moment directions and magnitudes are the result of inappropriate symmetry constraints, we also performed refinements using the triclinic magnetic space group P$_{\textrm{S}}1$ (\# 1.3). In this magnetic space group, both Nd sublattices are allowed to have moment components along all three crystallographic directions. However, in these refinements the $ab$-plane component for both sublattices was still constrained to lie along the [$1$ $1$ $0$] direction to facilitate comparison to the P$_{\textrm{b}}$m model. This refinement produced only a marginal improvement in the agreement factors and resulted in ordered moment magnitudes that were almost identical to those obtained for the P$_{\textrm{b}}$m model. In Table \ref{tab:MagneticRefinementResults} we show a comparison of the results from refinements with the two different magnetic space groups. 

As an additional approach to analyzing the data, we also employed the irep-based approach from representation analysis. This approach provides additional flexibility in addressing the robustness of key features of the magnetic space group approach solution from a fitting perspective. The representational analysis to determine the allowed irreps of the magnetic structure consistent with the magnetic propagation vector was conducted using the program BasIreps \cite{rodriguez2010program,rodriguez2021developments}. 

Due to the uncertainty regarding the nuclear structure, we performed refinements of the $T$ = 45 mK data using the ireps associated with the P6$_{3}$/m, P$\bar{3}$, and P3 space groups. In all cases, the associated ireps can be divided into two categories: those producing moments completely along the $c$-axis and those producing moments completely within the $ab$-plane. As discussed previously, the data indicate both $c$-axis and $ab$-plane moments, so the solution should involve a combination of ireps from both categories. Regardless of the space group chosen, refinements using only ireps that place the moments purely along the $c$-axis give poor agreement even when the (0 0 $L$)-type magnetic Bragg peaks are removed, reinforcing that there is a finite $ab$-plane component which is reflected in all peak intensities. 

No statistically satisfactory fits using the irreps of P6$_{3}$/m could be found, with all attempts yielding $R_f \geq$ 25 and poor agreement. When refining using the ireps of P$\bar{3}$ a significantly improved solution could only be accomplished by using a combination of three ireps, giving $R_{f}$ = 12.3. The refined magnetic structure showed two Nd magnetic sublattices with distinct moment directions and magnitudes. In the hexagonal coordinate system the two refined moments were [0.29(3) 0.29(3) 0.06(9)] $\mu_{B}$  and  [0.29(3) 0.29(3) 3.23(9)] $\mu_{B}$. Additional refinements in P3, where the two Nd sites are distinct at the crystallographic level, gave a very similar result as P$\bar{3}$.

To further investigate whether this type of solution was incorrectly favored due inappropriate symmetry constraints resulting from the uncertainty in the nuclear structure we refined the magnetic structure using the ireps of P1. This places no symmetry constraints on the moment directions of the two Nd sites (the direction of the $ab$-plane components was allowed to vary freely), only enforcing the effect of the $\vec{k}$ = (0 0 $\frac{1}{2}$) magnetic propagation vector. The converged solution possessed distinct Nd moments with a very similar arrangement to the best solutions found using P$\bar{3}$ or P3. To address the possibility that this solution represents a pernicious local minimum, we also tried a randomized brute force approach where the magnetic structure factors for a large number (50,000) of randomly chosen sets of basis vector coefficients for the best-fit irep combinations of the different space groups were calculated and compared to the data. The best goodness-of-fit solutions found in this way all showed the same unequal Nd moment behavior found through least-squares refinement. 

These analyses do not ameliorate the difficulties in physically reconciling the magnetic structure solution given in the Results section with the other available data. However, they do indicate that the feature of distinct moment magnitudes and directions is an aspect of the best fit solution regardless of the refinement method utilized. It is also not the result of inappropriate symmetry constraints or a local minimum in the goodness-of-fit parameter.

\begin{table*}[h!]
  \centering
  \setlength{\tabcolsep}{6pt}   
  \renewcommand{\arraystretch}{1.4} 
    \caption{Crystallographic description of the magnetic structure of \NC resulting from refinement of the $T$ = 45 mK data using two different magnetic space groups (MSG). A total of 44 integrated magnetic Bragg peaks were used for the refinement. The atomic positions (and their associated errors) are from the refinement of the nuclear structure in the P6$_{3}$/m spacegroup (see Table \ref{tab:NuclearRefinementResults}), which was performed separately from the magnetic refinement. Symmetry constrained atomic positions are given as fractions while free positions are decimal values. The magnetic moments are given in the hexagonal basis. }
\label{tab:MagneticRefinementResults}
\begin{tabular*}{18cm}{c | c | c}
\toprule
Parent space group & \multicolumn{2}{c}{ P6$_{3}$/m} \\
Propagation vector & \multicolumn{2}{c}{ (0 0 $\frac{1}{2}$)} \\
Transformation from the nuclear to magnetic unit cell & \multicolumn{2}{c}{($a$, $b$, 2$c$; 0, 0, 0)} \\
Unit cell parameters  & \multicolumn{2}{c}{$a$ = 7.414(6) \r{A}, $b$ = 7.414(6) \r{A}, c = 8.446(10) \r{A}} \\
& \multicolumn{2}{c}{$\alpha$ = 90\textdegree, $\beta$ = 90\textdegree, $\gamma$ = 120\textdegree} \\ \
Positions of magnetic atoms       & \multicolumn{2}{c}{Nd(1): $\frac{1}{3}$  $\frac{2}{3}$  $\frac{1}{8}$}  \\
                                  & \multicolumn{2}{c}{Nd(2): $\frac{2}{3}$  $\frac{1}{3}$  $\frac{3}{8}$}  \\ 
Position of non-magnetic atoms    & \multicolumn{2}{c}{Cl(1): 0.3868(6) 0.3007(6) $\frac{1}{8}$} \\
                                  & \multicolumn{2}{c}{Cl(2): 0.6993(6) 0.0861(6) $\frac{1}{8}$}  \\
                                  & \multicolumn{2}{c}{Cl(3): 0.9139(6) 0.6132(6) $\frac{1}{8}$} \\
                                  & \multicolumn{2}{c}{Cl(4): 0.6132(6) 0.6993(6) $\frac{3}{8}$ }  \\ 
                                  & \multicolumn{2}{c}{Cl(5): 0.3007(6) 0.9139(6) $\frac{3}{8}$} \\
                                  & \multicolumn{2}{c}{Cl(6): 0.0861(6) 0.3868(6)  $\frac{3}{8}$} \\ \hline
MSG symbol & P$_{\textrm{b}}$m& P$_{\textrm{S}}1$ \\
MSG number & 6.22 & 1.3 \\
MSG symmetry operations & 1. \textit{x, y, z}, +1                           & 1. \textit{x, y, z}, +1 \\
                        & 2. \textit{x, y, -z}+$\frac{3}{4}$, +1            &  \\
MSG symmetry centering operations & 1. \textit{x, y, z}, +1                 & 1. \textit{x, y, z}, +1 \\
                                  & 2. \textit{x, y, z}+$\frac{1}{2}$, -1   & 2. \textit{x, y, z}+$\frac{1}{2}$, -1\\
Magnetic moment components allowed by symmetry & Nd(1): [ $m_{a}$ $m_{b}$ $0$] & Nd(1): [ $m_{a}$ $m_{b}$ $m_{c}$] \\
Refined magnetic moment components & Nd(1): [0.60(5) 0.60(5) 0] $\mu_{B}$ & Nd(1): [0.61(5) 0.61(5) 0.02(10)] $\mu_{B}$ \\
                                    &Nd(2): [0 0 3.1(1)]$\mu_{B}$         & Nd(2): [0.01(20) 0.01(20) 3.1(1)] \\
Refined magnetic moment magnitudes          & Nd(1): 0.60(5) $\mu_{B}$             & Nd(1): 0.61(5) $\mu_{B}$ \\
                                    & Nd(2): 3.1(1) $\mu_{B}$           & Nd(2): 3.1(1) $\mu_{B}$ \\
    \bottomrule
  \end{tabular*}
\end{table*}

\FloatBarrier

%

\end{document}